\documentclass[conference]{IEEEtran}

\usepackage{cite,bbm,graphicx,amsmath,amssymb,mathrsfs,epsf} \usepackage{multirow}
\usepackage{subfigure,cite,graphicx,epsfig,amsmath,amssymb,mathrsfs,epsf,graphics,color,enumerate}

\newtheorem{theorem}{Theorem}
\newtheorem{lemma}{Lemma}
\newtheorem{example}{Example}
\newtheorem{remark}{Remark}

\newcommand{\uin}{}
\newcommand{\lin}{\bar}

\newcommand{\uset}{}
\newcommand{\lset}{\bar}

\begin{document}
\title{Noisy Network Coding with Partial DF}

\author{\authorblockN{Si-Hyeon Lee}
\authorblockA{Department of Electrical and Computer Engineering\\
University of Toronto, Toronto, Canada\\
Email: sihyeon.lee@utoronto.ca}
\and
\authorblockN{Sae-Young Chung}
\authorblockA{Department of Electrical Engineering \\
KAIST, Daejeon, Korea\\
Email: sychung@ee.kaist.ac.kr}}

\maketitle
\begin{abstract}
In this paper, we propose a noisy network coding integrated with partial decode-and-forward relaying for single-source multicast discrete memoryless networks (DMN's). Our coding scheme generalizes the partial-decode-compress-and-forward scheme (Theorem 7) by Cover and El Gamal. This is the first time the theorem is generalized for DMN's such that each relay performs both partial decode-and-forward and compress-and-forward simultaneously. Our coding scheme simultaneously generalizes both noisy network coding by Lim, Kim, El Gamal, and Chung and distributed decode-and-forward  by Lim, Kim, and Kim. It is not trivial to combine the two schemes because of inherent incompatibility in their encoding and decoding strategies. We solve this problem by sending the same long message over multiple blocks at the source and at the same time by letting the source find the auxiliary covering indices that carry information about the message simultaneously over all blocks. 
\end{abstract}

\IEEEpeerreviewmaketitle

\section{Introduction}

Decode-and-forward (DF) and compress-and-forward (CF) are the two main relaying strategies developed in~\cite{CoverElGamal:79} for the three-node relay channel. There have been many attempts to generalize them to discrete memoryless networks (DMN). For example, CF was generalized for DMN in~\cite{AvestimehrDiggaviTse:11,YassaeeAref:11,Lim:10} and DF was generalized for DMN in~\cite{KramerGastparGupta:05, LimKimKim:14}. For the three-node relay channel, the partial-decode-compress-and-forward relaying in Theorem 7 in~\cite{CoverElGamal:79} combines both DF and CF. However, to the best of our knowledge, there have been no results that generalize this to DMN. There were some relaying strategies for general DMN where each relay performs either CF or DF \cite{KramerGastparGupta:05,HouKramer:arxiv13}, but not both.

In this paper, we propose a unified scheme that generalizes the partial-decode-compress-and-forward relaying in Theorem 7 in~\cite{CoverElGamal:79} to DMN. Our scheme, which we call noisy network coding with partial DF (NNC-PDF), simultaneously generalizes noisy network coding (NNC) by Lim, Kim, El Gamal, and Chung~\cite{Lim:10} and distributed decode-and-forward (DDF) by Lim, Kim, and Kim~\cite{LimKimKim:14}. In our scheme, each relay can perform both partial DF and CF simultaneously.

It is not trivial to combine NNC and DDF schemes. In NNC, the same long message is sent over multiple blocks. In each block, the source sends an independent codeword indexed by the message and each relay compresses its received channel output sequence and forwards the covering index. After the transmission of all blocks, the destination decodes the message (uniquely) and the covering indices (nonuniquely) simultaneously over all blocks. On the other hand, in the DDF scheme, independent messages are sent over multiple blocks and information about each message is captured by a set of auxiliary covering indices, one per relay. Before starting the transmission, the source finds those auxiliary covering indices for each message using backward encoding over all blocks. In each block, the relay decodes the auxiliary covering index intended for itself and forwards it. After the transmission of all blocks, the destination decodes the message and the auxiliary covering indices sequentially using backward decoding.

Due to these inherent incompatibility in encoding and decoding strategies of NNC and DDF schemes, it is worthwhile to consider recent works on short message NNC \cite{YassaeeAref:11,HouKramer:arxiv13}. These works differ from \cite{LimKimKim:14} in that the source sends independent messages over blocks and the destination decodes the message using sliding window decoding \cite{YassaeeAref:11} or backward decoding \cite{HouKramer:arxiv13}, and hence more compatible with DDF scheme. One drawback of short message NNC schemes is that they require the destination to uniquely decode the covering indices, which results in extra constraints. These extra constraints turn out to be redundant \cite{YassaeeAref:11}, but when combined with partial DF, it is not clear whether similar arguments apply. This motivates us to consider a long message at the source and simultaneous nonunique decoding at the destination as in \cite{LimKimKim:14} for combining NNC and DDF schemes. In our NNC-PDF scheme, before starting the transmission, the source finds the auxiliary covering indices that carry information about the message simultaneously over all blocks (no backward encoding). In each block, each relay decodes the auxiliary covering index, compresses its received channel output sequence, and forwards the decoded and covered indices. After the transmission of all blocks, the destination decodes the message uniquely and the auxiliary covering indices and the covering indices nonuniquely, simultaneously over all blocks. 

The following notations are used throughout the paper.
For two integers $i$ and $j$, $[i:j]$ denotes the set $\{i,i+1,\ldots, j\}$. For a set $S$ of real numbers,   $S[i]$ denotes $\{j: j\in S, j<i\}$. For constants $u_1,\ldots, u_k$ and $S\subseteq [1:k]$, $u_S$  denotes the  vector $(u_{j}:j\in S)$ and $u^j_i$ denotes $u_{[i:j]}$ where the subscript is omitted when $i=1$, i.e., $u^j=u_{[1:j]}$.  For random variables $U_1,\ldots, U_k$ and $S\subseteq [1:k]$, $U_{S}$ and  $U^j_i$ are defined similarly. For sets $T_1, \ldots, T_k$ and $S\subseteq [1:k]$, $T_S$ denotes $\bigcup_{j\in S} T_j$ and $T^j_i$ denotes $T_{[i:j]}$ where the subscript is omitted when $i=1$. $\mathbbm{1}_{u=v}$ is the indicator function, i.e., it is 1 if $u=v$ and 0 otherwise. We follow the notion of typicality in \cite{Orlitsky:01}, \cite{ElGamalKim:11}.

\section{Model} \label{sec:model}
A single-source multicast discrete memoryless network of $N$ nodes $(\mathcal{X}_1,\ldots,\mathcal{X}_N, \mathcal{Y}_1,\ldots,\mathcal{Y}_N, p(y_{[1:N]}|x_{[1:N]}))$
consists of a set of channel input and output alphabets $(\mathcal{X}_k,\mathcal{Y}_k)$ for $k\in [1:N]$ and a collection of conditional probability mass functions $p(y_{[1:N]}|x_{[1:N]})$. 
Let node 1 denote the source node and let $\mathcal{D}\subseteq [2:N]$ denote the set of destination nodes. An $(R,n)$ code for the single-source multicast discrete memoryless network consists of message $I$, uniformly distributed over $\mathcal{I}\triangleq[1:2^{nR}]$, encoding function at the source that maps $I\in \mathcal{I}$ to $x_1^n\in \mathcal{X}_1^n$,  processing function at node $k\in [2:N]$ at time $i\in [1:n]$ that maps $y_k^{i-1}\in \mathcal{Y}_k^{i-1}$ to $x_{k,i}\in \mathcal{X}_{k}$, and decoding function at destination $d\in \mathcal{D}$ that maps $y_d^n\in \mathcal{Y}_d^n$ to $\hat{I}_d\in \mathcal{I}$. 
The probability of error is defined as $P_e^{(n)}=P(\hat{I}_d \neq I \mbox{ for some } d\in \mathcal{D})$ and a rate $R$ is said to be achievable if there exists a sequence of $(R,n)$ codes such that $\lim_{n\rightarrow \infty} P_e^{(n)}=0$. The capacity is defined as the supremum over all achievable rates.

\section{Noisy Network Coding with Partial Decode-and-Forward} \label{sec:main}
The following theorem gives a noisy network coding with partial decode-and-forward (NNC-PDF) bound for single-source multicast discrete memoryless networks, which simultaneously generalizes NNC \cite{Lim:10} and DDF \cite{LimKimKim:14}.

\begin{theorem} \label{thm:GDCF}
For a single-source multicast DMN, a rate of $R$ is achievable if  
\begin{align*}
R&<\min_{d\in \mathcal{D}}~\min_{S,T: S\subseteq T\subseteq [2:N]\setminus \{d\}} I(X_1,V_S;U_{S^c},X_{T^c},\hat{Y}_{T^c},Y_d|V_{S^c})\cr
&+I(X_{T},U_S;\hat{Y}_{T^c},Y_d|X_1, X_{T^c}, V^N, U_{S^c})\cr
&-I(\hat{Y}_{T};Y_{T}|\hat{Y}_{T^c}, X^N, V^N, U^N, Y_d)\cr
&-\sum_{k\in S^c} (I(U_k;X^N,V^N,U_{S^c[k]}|V_k,X_k,Y_k)+I(V_k;V_{S^c[k]}))
\end{align*}
for some $p(x_1,v_2^N,u_2^N)\prod_{k=2}^Np(x_k|v_k)p(\hat{y}_k|x_k,u_k,v_k,y_k)$ such that 
$\sum_{k\in S'}I(U_k;Y_k|X_k,V_k)>\sum_{k\in S'} I(V_k;V_{S'[k]})+I(U_k;U_{S'[k]},V_{S'}|V_k)$
for all $S'\subseteq [2:N]$.
\end{theorem}

\begin{remark}
For $N=3$, the NNC-PDF bound falls back to Theorem 7 in \cite{CoverElGamal:79}, which is a partial-decode-compress-and-forward bound for three node relay networks. 
\end{remark}

\begin{remark}
The NNC-PDF bound recovers NNC \cite{Lim:10} and DDF \cite{LimKimKim:14} bounds by letting $U^N=V^N=\emptyset$ and by letting $p(v_{[2:N]})=\prod_{k\in [2:N]}p(v_k)$, $V_k=X_k$, and $\hat{Y}_k=\emptyset$ for $k\in [2:N]$, respectively. 
\end{remark}

\begin{remark}
In the DDF scheme \cite{LimKimKim:14}, the relay nodes, i.e., nodes $2,\ldots,N-1$, send independent codewords. In contrast, the NNC-PDF scheme allows arbitrary correlation among transmitted codewords from relay nodes. 
\end{remark}

\section{Proof} 
Our NNC-PDF bound is derived as an application of our unified coding approach for network information theory~ \cite{LeeChung:arxiv14}, \cite{LeeChung:15ISIT1}. Let us briefly present the unified coding framework and unified coding theorem \cite{LeeChung:arxiv14}, \cite{LeeChung:15ISIT1} and then prove Theorem \ref{thm:GDCF}. 

\subsection{Unified approach for network information theory}
\subsubsection{Unified framework}
Let us first explain the unified framework \cite{LeeChung:arxiv14}, \cite{LeeChung:15ISIT1} for proving the achievability of many network information theory problems. An $N$-node acyclic discrete memoryless network (ADMN) $(\mathcal{X}_1,\ldots,\mathcal{X}_N$, $\mathcal{Y}_1,\ldots, \mathcal{Y}_N,\prod_{k=1}^Np(y_k|y^{k-1},x^{k-1}))$ consists of a set of alphabet pairs $(\mathcal{X}_k, \mathcal{Y}_k)$, $k\in [1:N]$ and a collection of conditional pmfs $p(y_k|y^{k-1},x^{k-1})$, $k\in [1:N]$. Here, $Y_k$ and $X_k$ represent any information that comes into and goes out of node $k$, respectively. 
In this network, information flows in one direction and node operations are sequential. Let $n$ denote the number of channel uses. First, $Y_1^n$ is generated according to $\prod_{i=1}^n p(y_{1,i})$ and then node 1 processes $X_1^n$ based on $Y_1^n$. Next, $Y_2^n$ is generated according to $\prod_{i=1}^n p(y_{2,i}|x_{1,i},y_{1,i})$ and then node 2 encodes $X_2^n$ based on $Y_2^n$. Similarly, $Y_k^n$ is generated according to $\prod_{i=1}^np(y_{k,i}|x^{k-1}_i,y^{k-1}_i)$ and node $k$ encodes $X_k^n$ based on $Y_k^n$ for $k\in [1:N]$. Clearly, any layered network \cite{AvestimehrDiggaviTse:11} or noncausal network (without infinite loop) \cite{ElGamalHassanpourMammen:07} possibly with noncausal state or side information can be represented as an ADMN. Furthermore, any strictly causal (usual discrete memoryless network with relay functions having one sample delay) or causal network (relays without delay \cite{ElGamalHassanpourMammen:07})  with blockwise operations can be represented as an ADMN by unfolding the network.

The objective of an ADMN is specified using a target joint distribution $p^*(x^N,y^N)$, which is shortly denoted as $p^*$. The probability of error is defined as $P_e^{(n)}(p^*, \epsilon)=P((X_{[1:N]}^n,Y_{[1:N]}^n) \notin \mathcal{T}_{\epsilon}^{(n)} )$, where the typical set $\mathcal{T}_{\epsilon}^{(n)}$ is defined with respect to $p^*$. We say the target distribution $p^*$ is achievable if there exists a sequence of node processing functions $Y_k^n\rightarrow X_k^n$, $k=1,\ldots, N$, such that $\lim_{n\rightarrow \infty} P_e^{(n)}(p^*,\epsilon)=0$ for any $\epsilon>0$. 

In the following, we illustrate how the point-to-point channel coding problem can be represented using an ADMN and a target distribution. 
\begin{example} For point-to-point channel coding problem with message rate of $R$, we choose $p(y_1)p(y_2|x_1,y_1)$ such that $H(Y_1)=R$ (corresp. to the message of rate $R$) and $p(y_2|x_1,y_1)=p(y_2|x_1)$ (corresp. to the channel) and pick $p^*$  such that $X_2$ (corresp. to the message estimate at node 2) is equal to $Y_1$. Then, $P_e^{(n)}(p^*, \epsilon)$ for any $\epsilon >0$ is lower-bounded by $P(Y_1^n\neq X_2^n)$, and hence achievability of $p^*$ implies that of rate $R$.
\end{example} 

\subsubsection{Unified coding theorem}
Let us present the unified coding theorem which gives a sufficient condition to achieve $p^*$ for an ADMN. To do that, we explain the coding parameters for the codebook generation and node operation of the unified coding scheme. In the unified coding scheme, $\nu$ covering codebooks $\mathcal{C}_1,\ldots,\mathcal{C}_{\nu}$ are generated to compress information that each node observes and decodes. Let $\mathcal{U}_j$ for $j\in [1:\nu]$ denote the alphabet for the codeword symbol of $\mathcal{C}_j$. For indexing of codewords, we consider $\mu$ index sets $\mathcal{L}_1,\ldots, \mathcal{L}_{\mu}$, where $\mathcal{L}_j=[1:2^{nr_j}]$ for some $r_j\geq 0$ for each $j\in [1:\mu]$.  We denote by $\Gamma_j\subseteq [1:\mu]$ the set of indices of $\mathcal{L}$'s associated with $\mathcal{C}_j$ in a way that each codeword in $\mathcal{C}_j$ is indexed by the vector $l_{\Gamma_j}\in \prod_{i\in \Gamma_j} \mathcal{L}_i$ and hence $\mathcal{C}_j$ consists of $2^{n\sum_{i\in \Gamma_j}r_i}$ codewords, i.e., $\mathcal{C}_j=\{u_j^n(l_{\Gamma_j}): l_{\Gamma_j}\in \prod_{i\in \Gamma_j} \mathcal{L}_i\}$.  
Each codebook is constructed allowing superposition coding. Let $A_j\subseteq [1:\nu], j\in [1:\nu]$ denote the set of the indices of $\mathcal{C}$'s on which $\mathcal{C}_j$ is constructed by superposition.

\begin{figure}[t]
 \centering
  {
   \includegraphics[width=90mm]{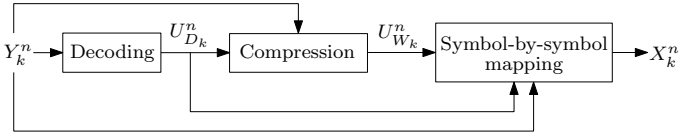}}
  \caption{Node $k\in [1:N]$ operates in three steps: 1) simultaneous nonunique decoding, 2) simultaneous compression, and 3) symbol-by-symbol mapping.} \label{fig:scheme_flow}
\end{figure}

 Node $k\in [1:N]$ operates according to the following three steps as illustrated in Fig. \ref{fig:scheme_flow}:  
 \begin{itemize}
\item Simultaneous nonunique decoding:  After receiving $y_k^n$, node $k$ decodes some covering codewords of previous nodes simultaneously, where some are decoded uniquely and the others are decoded non-uniquely. 
We denote by $D_k\subseteq [1:\nu]$ and $B_k\subseteq [1:\nu]$ the sets of the indices of $\mathcal{C}$'s whose codewords are decoded uniquely and non-uniquely, respectively, at node $k$. 

\item Simultaneous compression: After decoding, node $k$ finds covering codewords simultaneously according to a conditional pmf $p(u_{W_k}|u_{D_k}, y_k)$ that carry some information about received channel output sequence $y_k^n$ and uniquely decoded codewords $u_{D_k}^n$, where $W_k\subseteq [1:\nu]$ denotes the set of the indices of  $\mathcal{C}$'s  used for compression. 

\item Symbol-by-symbol mapping: After decoding and compression, node $k$ generates $x_k^n$ by a symbol-by-symbol mapping from uniquely decoded codewords $u_{D_k}^n$, covered codewords $u_{W_k}^n$, and received channel output sequence $y_k^n$. Let $x_k(u_{D_k}, u_{W_k}, y_k)$ denote the function used for symbol-by-symbol mapping. 
\end{itemize}

More rigorous explanations for the codebook generation and node operation of the unified coding scheme are provided in \cite{LeeChung:arxiv14}.

In summary, our scheme requires the following set $\omega$ of coding parameters, where some constraints are added to  make the aforementioned codebook generation and node operation proper:
\begin{enumerate}
\item positive integers $\mu$ and $\nu$
\item alphabets $\mathcal{U}_j, j\in [1:\nu]$
\item $\mu$-rate tuple $(r_1,\ldots, r_{\mu})$
\item sets $\uin{W}_k\subseteq [1:\nu]\setminus \uin{W}^{k-1}, \uin{D}_k\subseteq \uin{W}^{k-1}, \uin{B}_k\subseteq \uin{W}^{k-1}\setminus \uin{D}_k$, $\Gamma_j\subseteq [1:\mu]$, and $A_j\subseteq [1:\nu]$ for $k\in [1:N]$ and $j\in [1:\nu]$ that satisfy 
\begin{enumerate}[{A}-1]
\item $\Gamma_{\uin{W}_k}\setminus \Gamma_{\uin{D}_k}$'s are disjoint,
\item $\Gamma_{A_j}\subseteq \Gamma_j$ and  $j'<j$ if $j'\in A_j$,
\item $A_{\uin{W}_k}\subseteq \uin{W}_k\cup \uin{D}_k$, $A_{\uin{B}_k}\subseteq \uin{D}_k\cup \uin{B}_k$, and $A_{\uin{D}_k}\subseteq \uin{D}_k$.
\end{enumerate}
\item a set of conditional pmfs $p(u_{\uin{W}_k}|u_{\uin{D}_k},y_k)$ and functions $x_k(u_{\uin{D}_k},u_{\uin{W}_k},y_k)$ for $k\in [1:N]$ such that $p(x_{[1:N]},y_{[1:N]})$ induced by $\prod_{k=1}^N \!p(y_k|y^{k-1}\!,x^{k-1})p(u_{\uin{W}_k}|u_{\uin{D}_k}\!,y_k)\mathbbm{1}_{x_k=x_k(u_{\uin{D}_k}\!,u_{\uin{W}_k}\!,y_k)}$ is the same as the target distribution $p^*(x_{[1:N]},y_{[1:N]})$.
\end{enumerate}

The following theorem gives a sufficient condition for achievability using the aforementioned unified coding scheme. 
\begin{theorem}[Unified coding theorem \cite{LeeChung:arxiv14}, \cite{LeeChung:15ISIT1}] \label{thm:main}
For an  $N$-node ADMN, $p^*$ is achievable if there exists $\omega\in \Omega$ such that for $1\leq k \leq N$ 
\begin{align}
\sum_{j\in \lset{S}_k} r_j &<\sum_{j\in  \uset{S}_k} I(U_j;U_{ \uset{S}_k[j]\cup  \uset{S}_k^c},Y_k|U_{A_j}) \label{eqn:main_sk}\\ 
\sum_{j\in \lset{T}_k} r_j&>\sum_{j\in  \uset{T}_k} I(U_j;U_{ \uset{T}_k[j]\cup \uin{D}_k},Y_k|U_{A_j})\label{eqn:main_tk}
\end{align}
for all $\lset{S}_k\subseteq \lin{D}_k\cup \lin{B}_k$ such that $\lset{S}_k\cap \lin{D}_k\neq \emptyset$ and for all $\lset{T}_k\subseteq \lin{W}_k$ such that $\lset{T}_k\neq \emptyset$, where $\lin{D}_k\triangleq \Gamma_{\uin{D}_k}$, $\lin{B}_k\triangleq \Gamma_{\uin{B}_k}\setminus \Gamma_{\uin{D}_k}$, $\lin{W}_k\triangleq \Gamma_{\uin{W}_k}\setminus \Gamma_{\uin{D}_k}$, 
\begin{align*}
 \uset{S}_k&\triangleq \{j:j\in \uin{D}_k\cup \uin{B}_k, \Gamma_j\cap \lset{S}_k\neq \emptyset \}, \\
 \uset{T}_k&\triangleq \{j:j\in \uin{W}_k, \Gamma_j\cap (\lset{T}_k\cup \lin{D}_k)^c=\emptyset \}. 
\end{align*}
\end{theorem}

\begin{remark}
(\ref{eqn:main_sk}) and (\ref{eqn:main_tk}) for $k\in [1:N]$ are the conditions for successful simultaneous nonunique decoding and simultaneous compression, respectively, at node $k$. 
\end{remark}

\subsection{Proof of Theorem \ref{thm:GDCF}} 
To derive the NNC-PDF bound in Theorem \ref{thm:GDCF}, we first represent our network model as an ADMN and $p^*$ and then apply the unified coding theorem for a specific choice of coding parameter set $\omega$. 
\subsubsection{Corresponding ADMN and $p^*$ for our problem} 
We note that our network model itself cannot be represented as an ADMN. By assuming a block-wise operation at each node, however, the network can be unfolded and represented as an ADMN. 

Achievability uses $B$ transmission blocks\footnote{For initialization, we use additional $(N-1)^2$ blocks before the transmission of $B$ blocks, which will be explained later. }, each consisting of $n$ channel uses. Let $Y_{k,b}^n$ and $X_{k,b}^n$ for $k\in [1:N]$ and $b\in [1:B]$ denote the channel output and channel input sequences, respectively, at node $k$ at block $b$. At the end of block $b-1$, where $b\in [1:B]$, node $k\in [1:N]$ encodes what to transmit in block $b$, i.e., $X_{k,b}^n$, using previously received channel outputs up to block $b-1$, i.e., $Y_{k,[1:b-1]}^n$. Using this block-wise operation at each node, we  unfold the network. 
In the unfolded network, we have $(B+1)N$ nodes. The operation of node $(k,b), k\in [1:N], b\in [1:B]$ corresponds to that of node $k$ of the original network transmitting in block $b$ based on the received channel outputs up to block $b-1$ and the operation of node $(d,B+1), d\in \mathcal{D}$ corresponds to that of node $d$ of the original network that estimates the message based on the received channel outputs up to block $B$. Let $Y_{k,b}^{\mathrm{unf}}$ and $X_{k,b}^{\mathrm{unf}}$ denote the channel output and channel input at node $(k,b)$ of the unfolded network, respectively. A message generated at the source is regarded as the channel output at node $(1,1)$, i.e., $Y_{1,1}^{\mathrm{unf}}=M$ such that $H(M)=BR$, and the message estimate at destination $d\in \mathcal{D}$ is regarded as the channel input at node $(d,B+1)$, i.e., $\mathcal{X}_{d,B+1}^{\mathrm{unf}}=\mathcal{M}$.  To reflect the fact that node $(k,b+1)$ is originally the same node as node $(k,b)$, we assume that node $(k,b)$ has an orthogonal link of sufficiently large rate to node $(k,b+1)$. 
Hence, for $k\in [1:N]$ and $b\in [1:B]$, we let $Y_{k,b+1}^{\mathrm{unf}}=(X_{k,b}^{\mathrm{unf}}, Y_{k,b})$ and let $p(y_{k,b}|y_{[1:N],[1:b]}^{\mathrm{unf}},y_{[1:k-1],b+1}^{\mathrm{unf}},x_{[1:N],[1:b]}^{\mathrm{unf}},x_{[1:k-1],b+1}^{\mathrm{unf}})=p_{Y_k|Y_{[1:k-1]},X_{[1:N]}}(y_{k,b}|y_{[1:k-1],b},x_{[1:N],b})$ where $x_{k,b}^{\mathrm{unf}}=(x_{k,b}, z_{k,b})$ for $x_{k,b}\in \mathcal{X}_k$ and $z_{k,b}\in \mathcal{Z}_{k,b}$ for some $\mathcal{Z}_{k,b}$ with arbitrarily large cardinality. 

We assume a target joint distribution $p^*(x_{[1:N],[1:B+1]}^{\mathrm{unf}},y_{[1:N],[1:B+1]}^{\mathrm{unf}})$ such that $X_{d,B+1}^{\mathrm{unf}}=Y_{1,1}^{\mathrm{unf}}$ for all $d\in \mathcal{D}$. Note that the achievability of $p^*$ for the unfolded network implies the achievability of rate $R$ for the original network.  

\subsubsection{Application of unified coding theorem} 
Now, let us choose $\omega\in \Omega$ to obtain the NNC-PDF bound. In our NNC-PDF scheme, before starting the transmission, the source finds the auxiliary covering indices  that carry information about the message simultaneously over all blocks. At each block, node $k$ decodes the auxiliary covering index intended for it  and finds the covering index  that carries information about its received channel output. After all transmission, destination node decodes the message uniquely and the auxiliary covering indices and the covering indices nonuniquely, simultaneously over all blocks. This operation can be translated to the following parameter set for unified coding. 

Fix  $p_{X_1,V_2^N,U_2^N}\prod_{k=2}^Np_{X_k|V_k}p_{\hat{Y}_k|X_k,U_k,V_k,Y_k}$.  Let $\mu=2BN-B+N$ and $\nu=4BN-3B-N+2$. Consider a $\mu$-rate tuple $(r_0, r_{1,b}, r_{k,b'}, r'_{k,b''}: b\in [1:B], b'\in [0,B], b''\in [0,B-1], k\in [2:N])$. For notational convenience, let us index the codebook $\mathcal{C}$ by the auxiliary random variable used for its generation, i.e., if a codebook consists of $u^n(1),\ldots,u^n(2^{nr})$ generated conditionally independently according to $\prod_{i=1}^n p(u_i|v_i)$ for some $r\geq 0$ and $p(u|v)$, we denote the codebook by $\mathcal{C}_U$. In addition, we index the index set $\mathcal{L}$ in the following way: $\mathcal{L}_{l_0}=[1:2^{nr_0}], \mathcal{L}_{l_{1,b}}=[1:2^{nr_{1,b}}],  \mathcal{L}_{l_{k,b'}}=[1:2^{nr_{k,b'}}]$, and $\mathcal{L}_{l_{k,b''}'}=[1:2^{nr'_{k,b''}}]$ for $b\in [1:B], b'\in [0,B], b''\in [0,B-1]$, and $k\in [2:N]$. The remaining coding parameters associated with each node are given as follows:

\begin{itemize}
\item Node $(1,1)$: 
\begin{align*} 
&W_{1,1}=\{U_0, X_{1,b}, U_{k,b}, V_{k,b}, k\in [2:N], b\in [1:B]\}\\
&\Gamma_{U_0}= \{l_0\},~ \Gamma_{X_{1,b}}=\{l_0, l_{1,b}, l_{2, b-1}, \ldots, l_{N,b-1}\} \\
&\Gamma_{U_{k,b}}=\{l_{k,b},l_{k,b-1}\},~\Gamma_{V_{k,b}}=\{l_{k,b-1}\} \\
&A_{X_{1,b}}=\{V_{k,b}, k\in [2:N]\},~ A_{U_{k,b}}=\{V_{k,b}\}\\
&p(W_{1,1}|D_{1,1}, y_{1,1}^{\mathrm{unf}})=\mathbbm{1}_{u_0=y_{1,1}^{\mathrm{unf}}}\\
&~~~~\cdot \prod_{b\in [1:B]} p_{X_1,V_2^N,U_2^N}(x_{1,b},v_{2,b}\ldots v_{N,b}, u_{2,b},\ldots, u_{N,b}) 
\end{align*}

\item Node $(k,1)$, $k\in [2:N]$: 
\begin{align*}
D_{k,1}=\{V_{k,1}\},&~ W_{k,1}=\{X_{k,1}\} \\
\Gamma_{X_{k,1}}=\{l_{k,0}, l'_{k,0}\},&~A_{X_{k,1}}=\{V_{k,1}\} \\
p(W_{k,1}|D_{k,1}, y_{k,1}^{\mathrm{unf}})&=p_{X_k|V_k}(x_{k,1}|v_{k,1}) 
\end{align*}


\item Node $(1,b)$, $b\in [2:B]$: 
\begin{align*}
D_{1,b}&=W_{1,1} 
\end{align*}

\item Node $(k,b)$, $k\in [2:N], b\in [2:B]$:
\begin{align*}
&D_{k,b}=W_{k}^{b-1}\cup D_k^{b-1}\cup \{U_{k,b-1}, V_{k,b}\} \\
&W_{k,b}=\{\hat{Y}_{k,b-1}, X_{k,b}\}  \\
&\Gamma_{\hat{Y}_{k,b-1}}=\{l'_{k,b-1},l_{k,b-1},l'_{k,b-2},l_{k,b-2}\}\\
&\Gamma_{X_{k,b}}=\{l_{k,b-1},l_{k,b-1}'\} \\
&A_{\hat{Y}_{k,b-1}}=\{X_{k,b-1}, U_{k,b-1}, V_{k,b-1}\},~ A_{X_{k,b}}=\{V_{k,b}\} \\
&p(W_{k,b}|D_{k,b}, y_{k,b}^{\mathrm{unf}})\\
&=p_{\hat{Y}_k|X_k,U_k,V_k,Y_k}(\hat{y}_{k,b-1}|x_{k,b-1},u_{k,b-1},v_{k,b-1},y_{k,b-1})\\
&~~~~~~\cdot p_{X_k|V_k}(x_{k,b}|v_{k,b}) 
\end{align*}

\item Node $(d,B+1)$, $d\in \mathcal{D}$
\begin{align*}
D_{d,B+1}&=W_{d}^{B}\cup D_d^{B}\cup \{U_0\} \\
B_{d,B+1}&= \{X_{1,b}, U_{k,b},  V_{k,b}, X_{k,b}, \hat{Y}_{k,b}, \\
&~~~~~~~~~~~~~~~k\in [2:N]\setminus \{d\}, b\in [1:B-1] \} \\
\end{align*}

\end{itemize}
For $k\in [1:N]$ and $b\in [1:B]$, we let $X_{k,b}^{\mathrm{unf}}=(Y_{k,[1:b-1]}, W_{k}^b, D_{k}^b)$. 
For $d\in D$, let $X_{d,B+1}^{\mathrm{unf}}=U_0$. Before initiating the transmission of $B$ blocks, each $l_{k,0}$ chosen at the source is transmitted to node $k\in [2:N]$ using multi-hop scheme as in \cite{HouKramer:arxiv13}, and this requires additional $(N-1)^2$ blocks. Note the actual rate $\frac{nBR}{nB+n(N-1)^2}$ can be arbitrarily close to $R$ as $B\rightarrow \infty$.  As a result of this initialization step, we can let $Y_{k,1}^{\mathrm{unf}}=V_{k,1}$ for $k\in [2:N]$. 

\begin{remark}
In the above choice of $\omega\in \Omega$, $l_0$ corresponds to the message index and $\{l_{k,b}:k\in [2:N], b\in [0:B]\}$ corresponds to the set of auxiliary covering indices that carry information about the message. These indices are chosen simultaneously at node (1,1) of the unfolded network, i.e., the source before starting the transmission. Node $(k,b)$ for $k\in [2:N]$, $b\in [2:B]$ of the unfolded network, i.e., node $k$ at the end of block $b-1$, decodes $l_{k,b-1}$ and finds the covering index $l_{k,b-1}'$ that carries information about $Y_{k,b-1}^n$. At node $(d,B+1)$ for $d\in \mathcal{D}$ of the unfolded network, i.e., node $d$ at the end of block $B$,  simultaneously decodes the message (uniquely) and the auxiliary covering indices and the covering indices (nonuniquely). 
\end{remark}

To prove Theorem \ref{thm:GDCF}, we use the following lemma, whose proof is given in \cite{LeeChung:arxiv14}.
\begin{lemma} \label{corollary:GDCF_reduced}
Consider $\omega\in \Omega$. For $\lset{S}_k\subseteq \lin{D}_k\cup \lin{B}_k$ such that $\lset{S}_k\cap \lin{D}_k\neq \emptyset$ and $\lset{T}_k\subseteq \lin{W}_k$ such that $\lset{T}_k\neq \emptyset$, the decoding and compression bounds, i.e., (\ref{eqn:main_sk}) and (\ref{eqn:main_tk}), in Theorem \ref{thm:main} are satisfied if 
\begin{align}
\sum_{j\in \bar{S}_k}r_j<\sum_{j\in S_k'}I(U_j;U_{S_k'[j]\cup S_k'^c},Y_k|U_{A_j})\\
\sum_{j\in \bar{T}_k}r_j>\sum_{j\in T_k'}I(U_j;U_{T_k'[j]\cup D_k},Y_k|U_{A_j}),
\end{align}
for some $S_k'\subseteq S_k$ such that $A_j\subseteq (S_k\setminus S_k')[j]\cup S_k^c$ for all $j\in S_k\setminus S_k'$ and $T_k\subseteq T_k'$.
\end{lemma}

Now, we are ready to apply Theorem \ref{thm:main} to prove Theorem \ref{thm:GDCF}. Let $r_{1,b}=r_1$, $r_{k,B}=0$, $r_{k,b'}=r_k$, and $r'_{k,b'}=r_k'$ for some $r_1\geq 0, r_k\geq 0$, and $r_k'\geq 0$ for $b\in [1:B]$, $b'\in [0:B-1]$, and $k\in [2:N]$. For each node in the unfolded network, let us derive the decoding and compression bounds i.e., (\ref{eqn:main_sk}) and (\ref{eqn:main_tk}), respectively. First, for compression at node (1,1), note that $\bar{W}_{1,1}=\{l_0,l_{1,b}, l_{k,b'}, b\in [1:B], b'\in [0:B], k\in [2:N]\}$. By applying Lemma \ref{corollary:GDCF_reduced} and using the following blockwise i.i.d. property $p(x_{1,[1:B]},v_{[2:N],[1:B]}, u_{[2:N],[1:B]}) =\prod_{b\in [1:B]}  p_{X_1,V_2^N,U_2^N}(x_{1,b},v_{[2:N],b}, u_{[2:N],b})$, we can show that the condition for compression is satisfied if
\begin{align*}
r_0&>BR \\
\sum_{k\in S} r_{k} &> \sum_{k\in S}I(V_{k};V_{S[k]}) +\sum_{k\in S} I(U_{k};U_{S[k]},V_{S}|V_{k})\\
r_1+\sum_{k\in [2:N]} r_{k} &> \sum_{k\in [2:N]}I(V_{k};V^{k-1}) \cr
&+\sum_{k\in [2:N]} I(U_{k};U^{k-1},V_{[2:N]},X_1|V_{k})
\end{align*} 
for all $S\subseteq [2:N]$. 

Next, consider the decoding and compression bounds for node $(k,1), k\in [2:N]$. Since $Y_{k,1}^{\mathrm{unf}}=D_{k,1}$, the bound for decoding  becomes inactive. Furthermore, it can be easily shown that the bound for compression is also inactive.  Similarly, for node $(1,b), b\in [2:B]$, since $Y_{1,b}^{\mathrm{unf}}=D_{1,b}$, the bound for decoding becomes inactive.

Now, consider the decoding and compression bounds at node $(k,b), k\in [2:N], b\in [2:B]$. Since $\bar{D}_{k,b}=\bar{W}_{k,b-1}\cup \bar{D}_{k,b-1} \cup \{l_{k,b-1}\}$ and $Y_{k,b}^{\mathrm{unf}}$ contains $W_{k,b-1}$ and $D_{k,b-1}$, we only need to consider $\bar{S}_{k,b}=\{l_{k,b-1}\}$\footnote{We can show rigorously that bounds corresponding to $\bar{S}_{k,b}$ such that $\bar{S}_{k,b}\cap (\bar{W}_{k,b-1}\cup \bar{D}_{k,b-1})\neq \emptyset$ become redundant, but we omit the proof because it is trivial.}.  From the following blockwise i.i.d. property 
\begin{align}
& p(x_{[1:N],[1:B-1]},\!v_{[2:N],[1:B-1]}, \!u_{[2:N],[1:B-1]}, \!y_{[1:N]},\hat{y}_{[2:N],[1:B-1]}) \cr
&=\prod_{b\in [1:B-1]} \big( p_{X_1,V_2^N,U_2^N}(x_{1,b},v_{[2:N],b}, u_{[2:N],b})\cr
&~~~~\cdot \prod_{k\in [2:N]} p_{X_k|V_k}(x_{k,b}|v_{k,b}) \cdot p_{Y_{[1:N]}|X_{[1:N]}}(y_{[1:N],b}|x_{[1:N],b})\cr 
&~~~~\cdot \prod_{k\in [2:N]} p_{\hat{Y}_k|X_k,U_k,V_k,Y_k}(\hat{y}_{k,b}|x_{k,b},u_{k,b},v_{k,b},y_{k,b})\big), \label{eqn:GDCF_independence}
\end{align}
the bound for decoding is given as  $r_{k}<I(U_k;Y_k|X_k,V_k)$. Also, we have $\bar{W}_{k,b}=\{l_{k,b-1}'\}$. From the blockwise i.i.d. property shown in (\ref{eqn:GDCF_independence}), the bound for compression is given as $
r_{k}'>I(\hat{Y}_k;Y_k|X_k,U_k,V_k)$.

Lastly, consider the decoding bound at node $(d,B+1), d\in \mathcal{D}$. Since  $\bar{D}_{d,B+1}=\bar{W}_{d,B}\cup \bar{D}_{d,B} \cup \{l_0\}$, where $\bar{W}_{d,B}\cup \bar{D}_{d,B} =\{l_{d,b}, l_{d,b}': b\in [0:B-1]\}$, and $Y_{d,B+1}^{\mathrm{unf}}$ contains $W_{d,B}$ and $D_{d,B}$, we only need to consider $\bar{S}_{d,B+1}\subseteq \{l_0, l_{1,b}, l_{k,b'}, l_{k,b'}': b\in [1:B-1], b'\in [0:B-1], k\in [2:N]\setminus \{d\}\}$ such that $l_0\in \bar{S}_{d,B+1}$. From Lemma \ref{corollary:GDCF_reduced} and using the blockwise i.i.d. property shown in (\ref{eqn:GDCF_independence}), we can show that the bound for decoding is satisfied if 
\begin{align*}
&r_0<(B-1)\cr
&\times\min_{S,T: S\subseteq T\subseteq [2:N]\setminus \{d\}} \Big(\sum_{k\in S} I(V_k;V_{S[k]}, V_{S^c}, U_{S^c}, \hat{Y}_{T^c}, X_{T^c}, Y_d) \cr
&+I(X_1;U_{S^c}, \hat{Y}_{T^c}, X_{T^c}, Y_d|V_{[2:N]})\cr
&+\sum_{k\in T} I(X_k;X_{T[k]}, X_1, V_{[2:N]},U_{S^c}, \hat{Y}_{T^c}, X_{T^c}, Y_d|V_k) \cr
&+\sum_{k\in S} I(U_k;U_{S[k]}, X^N,  V_{[2:N]},U_{S^c}, \hat{Y}_{T^c},Y_d|V_k) \cr
&+\sum_{k\in T} I(\hat{Y}_k;\hat{Y}_{T[k]}, \hat{Y}_{T^c},X^N,  V_{[2:N]},U_{[2:N]}, Y_d|U_k,V_k,X_k) \cr
&-(\sum_{k \in S}r_{k} +\sum_{k\in T}r_{k}')-r_1\Big)-\sum_{k\in [2:N]} (r_k+r_k').
\end{align*}


By performing Fourier-Motzkin elimination and taking $B\rightarrow \infty$, $p^*$ is shown to be achievable if the condition in Thoerem \ref{thm:GDCF} is satisfied, and hence Theorem \ref{thm:GDCF} is proved. 

\section*{Acknowledgement}
This work was supported in part by the CISS funded by the Ministry of Science, ICT \& Future Planning as the Global Frontier Project.”


\begin{thebibliography}{10}
\providecommand{\url}[1]{#1}
\csname url@samestyle\endcsname
\providecommand{\newblock}{\relax}
\providecommand{\bibinfo}[2]{#2}
\providecommand{\BIBentrySTDinterwordspacing}{\spaceskip=0pt\relax}
\providecommand{\BIBentryALTinterwordstretchfactor}{4}
\providecommand{\BIBentryALTinterwordspacing}{\spaceskip=\fontdimen2\font plus
\BIBentryALTinterwordstretchfactor\fontdimen3\font minus
  \fontdimen4\font\relax}
\providecommand{\BIBforeignlanguage}[2]{{%
\expandafter\ifx\csname l@#1\endcsname\relax
\typeout{** WARNING: IEEEtran.bst: No hyphenation pattern has been}%
\typeout{** loaded for the language `#1'. Using the pattern for}%
\typeout{** the default language instead.}%
\else
\language=\csname l@#1\endcsname
\fi
#2}}
\providecommand{\BIBdecl}{\relax}
\BIBdecl

\bibitem{CoverElGamal:79}
T.~M. Cover and A.~{El Gamal}, ``Capacity theorems for the relay channel,''
  \emph{{IEEE} Trans. Inf. Theory}, vol.~25, pp. 572--584, Sep. 1979.

\bibitem{AvestimehrDiggaviTse:11}
A.~S. Avestimehr, S.~N. Diggavi, and D.~Tse, ``Wireless network information
  flow: {A} deterministic approach,'' \emph{{IEEE} Trans. Inf. Theory},
  vol.~57, pp. 1872--1905, April 2011.

\bibitem{YassaeeAref:11}
M.~H. Yassaee and M.~R. Aref, ``Slepian-{W}olf coding over cooperative relay
  networks,'' \emph{{IEEE} Trans. Inf. Theory}, vol.~57, pp. 3462--3482, Jun.
  2011.

\bibitem{Lim:10}
S.~H. Lim, Y.-H. Kim, A.~{El Gamal}, and S.-Y. Chung, ``Noisy network coding,''
  \emph{{IEEE} Trans. Inf. Theory}, vol.~57, pp. 3132--3152, May 2011.

\bibitem{KramerGastparGupta:05}
G.~Kramer, M.~Gastpar, and P.~Gupta, ``Cooperative strategies and capacity
  theorems for relay networks,'' \emph{{IEEE} Trans. Inf. Theory}, vol.~51, pp.
  3037--3063, Sep. 2005.

\bibitem{LimKimKim:14}
S.~H. Lim, K.~T. Kim, and Y.-H. Kim, ``Distributed decode--forward for
  multicast,'' in \emph{Proc. {IEEE} Int. Symp. Inform. Theory (ISIT)},
  Jun.-Jul. 2014, pp. 636--640.

\bibitem{HouKramer:arxiv13}
J.~Hou and G.~Kramer, ``Short message noisy network coding with a
  decode-forward option,'' \emph{{IEEE} Trans. Inf. Theory}, submitted for
  publication. [Online]. Available: http://arxiv.org/abs/1304.1692.

\bibitem{Orlitsky:01}
A.~Orlitsky and J.~R. Roche, ``Coding for computing,'' \emph{{IEEE} Trans. Inf.
  Theory}, vol.~47, pp. 903--917, Mar. 2001.

\bibitem{ElGamalKim:11}
A.~{El Gamal} and Y.-H. Kim, \emph{Network information theory}.\hskip 1em plus
  0.5em minus 0.4em\relax Cambridge, U.K.: Cambridge Univ. Press, 2011.

\bibitem{LeeChung:arxiv14}
S.-H. Lee and S.-Y. Chung, ``A unified approach for network information
  theory,'' [Online]. Available: http://arxiv.org/abs/1401.6023.

\bibitem{LeeChung:15ISIT1}
------, ``A unified approach for network information theory,'' \emph{\emph{to
  appear in }Proc. {IEEE} Int. Symp. Inform. Theory (ISIT)}, June 2015.

\bibitem{ElGamalHassanpourMammen:07}
A.~{El Gamal}, N.~Hassanpour, and J.~Mammen, ``Relay networks with delays,''
  \emph{{IEEE} Trans. Inf. Theory}, vol.~53, pp. 3413--3431, Oct. 2007.

\end{thebibliography}
\end{document}